# Cyber Warfare: Terms, Issues, Laws and Controversies

Kamile Nur Seviş
Cyber Security Engineering
Istanbul Sehir University
Istanbul TURKEY
nursevis@sehir.std.edu.tr

Ensar Seker
Cyber Security Institute
TUBITAK (The Scientific and Technological Research Council of Turkey)
Kocaeli, TURKEY
ensar.seker@tubitak.gov.tr

*Abstract* — **Recent years have shown us the importance of cybersecurity. Especially, when the matter is national security, it is even more essential and crucial. Increasing cyber attacks, especially between countries in governmental level, created a new term cyber warfare. Creating some rules and regulations for this kind of war is necessary therefore international justice systems are working on it continuously. In this paper, we mentioned fundamental terms of cybersecurity, cyber capabilities of some countries, some important cyber attacks in near past, and finally, globally applied cyber warfare law for this attacks.**

*Keywords-component; Law, National Security, Self-Defense Cyber Attack, Cyber Warfare.*

## I. Introduction

Since the technology is being developed incredibly fast, it certainly becomes part of our daily life and effects it either in a good way or in a bad way. It is fair to say that the internet became a central factor of these technological developments. The new phase of international relationships between states and domestic relationships between states with their public took place based on this modern technology. Instead of traditional ways, many states begun to use new communication methods and replaced the old ones. They now provide services to their citizens over a network such as passport and visa applications, tax and bill payments, exchange and open market operations, banking and insurance transactions, registration procedures for military services etc. All these services can be used via computers, tablets and even smart phones with an internet connection. Even though, there are many benefits of this system, every action has an equal and opposite reaction. Cyber terrorism, cyber threats, cyber espionage, and cyber war are the other side of the coin. While personal privacy was in danger against these perils, risks are much higher at national security level. Attackers are trying to steal, damage, destroy or control critical information from people, organizations, companies and even governments. Cybercrimes in a lower scale and cyber warfare in a higher scale easily can turn a daydream into a nightmare. Albeit these technological systems are getting more sophisticated, attacks against these structures are getting more complicated as well. Battlefields are now shifting from actual places to virtual areas. Besides their land forces, air forces and navy, countries create cyber armies nowadays. There are some certain applied rules and laws to use these forces and they need to be adapted to cyber war as well. Although countries have different types of laws for cyberspace domestically, there isn't any binding law for cyber warfare globally. This study focuses the laws, regulations and manuals which are developed for cyber warfare. Describe the general perspective of the chapter. Toward the end, specifically state the objectives of the chapter.

## II. CYBER TERMINOLOGY

### A. Cyberspace

The expression *cyberspace* initially showed up in a short story named "Burning Chrome" in the 1982 written by American author William Gibson and later in his 1984 novel "Neuromancer". In the following couple of years, the word turned out to be conspicuously related to online PC systems. The term *cyberspace* which, is used mostly to express the internet contains numerical interaction and communication method. According to the definition of White House "Cyberspace is composed of hundreds of thousands of interconnected computers, servers, routers, switches, and fiber optic cables that allow our critical infrastructures to work [1]."U.S. Department of Defense describes cyberspace as "A global domain within the information environment consisting of the interdependent network of information technology infrastructures and resident data, including the Internet, telecommunications networks, computer systems, and embedded processors and controllers [2]." NATO says, on the other hand, people are part of cyberspace as well and defines it as

"Cyberspace is more than the internet, including not only hardware, software and information systems, but also people and social interaction within these networks [3]."

*B. Cybersecurity*

Protecting frameworks, systems and information in cyberspace against cyber threats needs some sort of security. All tools, methods, guidance, activities, and technologies to defend assets and privacy of users, organizations, agencies and governments called cybersecurity. Main purpose of cybersecurity is preventing possible security hazards to assure safety and privacy of crucial data in the internet.

*C. Cyber Warfare*

Before all those technologies, there were only ground and maritime warfare. In 1900s, developments on aeronautic make aerial combat another battlefield. Space became a new warfare between super powers in 1950s. In the 21st century, cyber warfare joined to war terminology and described as a fifth combat zone. Even though there are many definitions for cyber warfare since there isn't any peace treaty for it, it's hard to describe which malicious activities are considered as cyber warfare. The most critical uncertainty is the edge for seeing a digital episode as the use of power. The privilege to self-preservation is activated by the use of power. This makes the subject of the edge between a demonstration that legitimizes the use of power accordingly (a demonstration of war) and a demonstration that does not key to the examination of cyber warfare. An act of war is the risk or use of power against the territorial integrity or political freedom [4].

By general definition "cyber warfare refers to a massively coordinated digital assault on a government by another, or by large groups of citizens. It is the action by a nation-state to penetrate another nation's computers and networks for the purposes of causing damage or disruption." Be that as it may, it includes that "the term cyber warfare may also be used to describe attacks between corporations, from terrorist organizations, or simply attacks by individuals called hackers, who are perceived as being warlike in their intent [2] [5]."

According to RAND corporation not just attacks even, attempts to harm another country's computers or data networks through are also considered as cyber warfare [6]. Including Advanced Persistent Threats (APT).

Richard Clarke defines cyberwar as "actions by a nation - state to penetrate another nation's computer or networks for the purposes of causing damage or disruption [7]." Based on his definition, this action should be between states and there should be damage to be cyberwar.

When we look at all definitions of cyber warfare, even though there is no reciprocal conflict one another if there are damage(s) and harmful results because of cyber attacks it can be characterized as cyberwar. Even if those virtual attacks didn't create impacts like real battles did so far, when we think about worst case scenarios there are extremely serious risks. Richard Clarke expresses how cyber warfare can be devastating like a real war for many countries, despite the fact it would be begun between two nations. He emphasizes that a cyberwar could begin and end so quickly, and systems which are used for traditional war can be destroyed easily by cyber attacks [7]. If we compare cyberwar with traditional wars, there are some main differences. One of them is the harm they create. It's hard to predict that what would be the damage after cyber attacks. Planning and executing cyber assaults might be in a short pass of time. When the system is affected, whole structure will fall rapidly. After the attackers infiltrate systems and gain the control, anything could happen depend on what kind of system was captured. Physical damages, injuries and even deaths could be the possible outcomes due to targeting physical systems which are controlled by computers. In case of extremely critical infrastructures are invaded by these cyber attacks such as a nuclear reactor, massive harm can be created and that would be catastrophic. Another difference is weapons. Weapons which are used for cyber battles such as computer systems, hardware and software are much cheaper and more available than real ammunitions. Determining the origin of the cyber attack and identifying the attackers aren't easy and might be the most important difference between cyberwar with a typical war. Hackers are incognito. They are nameless and they are faceless. Most of the time they don't have any identification or they use fake ones. Malicious activities steered through numerous regions. As long as the attackers accept what they did, it's almost impossible to provide hard proof for the felony. Finally, it is certainly more critical information can be gathered and more enormous damage can be caused with reasonable costs via cyber attacks. That's why it is understandable how cyberspace is getting to be center of the attention for the future. In the next chapter to get a better idea, cyber capabilities of some major countries and some important cyber attacks in near past will be studied.

### III. CYBER CAPABILITIES BY COUNTRY

Many nations have or are creating apparatuses for PC undercover work and assault including espionage and sabotage. While activities in cyberspace are military doctrine for some countries, for some others, they are part of their national security program. In his book, according to R. Clarke, a capability of countries on cyberspace varies based on their cyber dependencies, cyber attacks, and cyber defense [7]. United States, China, Russia and North Korea are the main actors on cyberspace.

*A. United States*

In 2002, first step was taken for cyber warfare strategies by the President. There are four main actors besides many other agencies which are USCYBERCOM "to plan, coordinate, integrate, synchronize and conduct activities to: direct the operations and defense of specified Department of

Defense information networks and; prepare to, and when directed, conduct full spectrum military cyberspace operations in order to enable actions in all domains, ensure US/Allied freedom of action in cyberspace and deny the same to our adversaries [8]," NSA/CSS (The National Security Agency/Central Security Service) "to lead the U.S. Government in cryptology that encompasses both Signals Intelligence (SIGINT) and Information Assurance (IA) products and services, and enables Computer Network Operations (CNO) in order to gain a decision advantage for the Nation and our allies under all circumstances [9]", US-CERT (The Department of Homeland Security's United States Computer Emergency Readiness Team) "to lead efforts to improve the Nation's cybersecurity posture, coordinate cyber information sharing, and proactively manage cyber risks to the Nation...[10]", and the FBI (Federal Bureau of Investigation) "to prevent harm to national security as the nation's domestic intelligence agency and to enforce federal laws as the nation's principal law enforcement agency [11]."

*B. China*

According to the annual report of the National Computer Network Emergency Response Technical Team Coordination Center of China (CNCERT or CNCERT/CC), China has noticeable IT infrastructure and advanced cyber weapons. China plays an active role on cyberspace and cyber attacks. Digital attacks against China and from China significantly increased in recent years [12]. A spokesman from China's Ministry of Defence and the People's Liberation Army (PLA) says cyberwar and cyber attacks are serious and important as much as other wars [13]. The announcement was made by National Defense of the People's Republic of China about establishment of Information Protection Base under the General Staff Department on 20 July 2010 [14]. In mid-2014, first professional "blue army" troop unit was built up by the PLA. The Blue Army offends data and a network when contrasted with conventional military [15].

*C. Russia*

When the internet is concerned, Russia is the most agonized over the risk postured by antagonistic. "The Federal Security Service (FSB) is a federal executive body with the authority to implement government policy in the national security of the Russian Federation ... ensuring the information security of Russia and exercising the basic functions of the federal security services specified in the Russian legislation ... [16] [17]." Own-created and off-the-track hacking devices utilized by FSB intelligence agency, which operates locally and globally. On the other hand, the Special Communications and Information Service of the Federal Protective Service of the Russian Federation (Spetssvyaz) and Federal Agency of Government Communications and Information (FAGCI/FAPSI) are in charge of the gathering and investigation of international communications and signal intelligence, and in addition, ensuring Russian government correspondences and data systems, which includes data security and cryptanalysis [18]. There are critical differentiations between western and Russian ideas of digital security. Russia has a statist idea of who ought to be included in the internet; Russian authorities affirm the guideline of national limits. Furthermore, the Russian idea of *breach of data space* is not popular in the West. Russia wants to create new international arrangements about cyberspace with support of its allies like China, Tajikistan and Uzbekistan [19].

*D. North Korea*

North Korea has more than 5,000 hacker forces based on South Korean's reports [20]. A new unit was initially established by North Korean military in 1998 that concentrates exclusively on digital warfare. The unit, named Unit 121 has consistently developed in size and ability from that point forward [21]. They have threatened everything from further ballistic rocket tests, another atomic test, withdrawal from the 1953 cease-fire that ended Korean War dangers (there is no peace settlement) and cyber warfare is no exception [22].

IV. MOST THREATING CYBER WARFARE ATTACKS IN NEAR PAST

Cyber attacks became much more sophisticated and complex since Robert Morris created the first computer worm to test the size of the internet in 1989 [23]. It is possible to say that cyberwar is already happening on cyberspace nowadays. Discovery of *Stuxnet* (the first cyber warfare weapon ever known) was a defining moment ever in the history of cybersecurity. As opposed to starting conviction, *Stuxnet* wasn't about mechanical undercover work. It didn't take, control, or delete data. Instead, *Stuxnet*'s objective was to physically annihilate a military target allegorically, as well as truly [24]. *Stuxnet* has obviously contaminated more than 60,000 PCs around the world, mostly in Iran [25]. A definitive objective of *Stuxnet* was to harm that facility by reprogramming programmable logic controllers (PLCs) to work as the attackers intend them to, no doubt out of their predefined limits [26]. In spite of the fact that the creators of *Stuxnet* haven't been formally recognized, the size and complexity of the worm have persuaded that it could have been made with support of state(s). Even though there isn't any conclusive evidence, it is believed that the United States and Israel behind *Stuxnet* [27].

Another country who suffered from cyber attacks was Estonia. Estonia's network infrastructure was targeted by hackers on April 27, 2007 and running for a time of a few weeks. No less than 128 exceptional DDOS assaults focusing on internet conventions in Estonia occurred amid this period. Internet traffic expanded from 20,000 packets to more than 4 million packets for every second [28]. Estonian authorities like Foreign Minister Urmas Paet immediately blamed Russia for executing the assaults, however European Commission and NATO specialized specialists were not able find enough evidence for confirmation of Kremlin participation in these cyber attacks. Following quite a long while of lobbying, Estonia as of late got NATO emergency courses of action to secure the nation in the case of a theoretical Russian intrusion [29].

Tanks, artillery and warplanes weren't the only weapons Russia used against the Georgia when the Russian-Georgian War begun in August of 2008. Before any gun was fired, cyber attacks were already hitting Georgia. 54 websites in Georgia related to communications, finance, and the government was targeted immediately. All communication channels were hacked for preventing Georgian public to be informed [30].

China has developed a communications intelligence program called *Golden Shield* using new and advanced technologies to gather domestic and foreign intelligence. There are too many state-level cyber attacks from China to United States so those attacks are even named. *Titan Rain* is a U.S. code name for Chinese military cyber assaults against the U.S. [31]. One of these attacks called *Operation Aurora*. Operation Aurora was a progression of digital assaults directed by advanced persistent threats (APTs) with binds to the PLA. The assault was to a great degree wide-scale and is believed to have focused on 34 organizations along with Yahoo, Symantec, Northrop Grumman, Morgan Stanley, Dow Chemical and Google. Sophistication of Operation Aurora makes the security vendor McAfee to believe it was produced by defense industry [32] [33].

In October 2012, Kaspersky Lab's group of specialists started an examination taking after a progression of assaults against computer networks targeting international diplomatic service agencies. A substantial scale cyber espionage activities system was uncovered and broke down amid the examination. It was named *Operation Red October*, called *Rocra* for short [34].

According to security experts, a complex targeted on digital assault that gathered private information from nations, for example, Israel and Iran has been revealed which known as *Flame*, had been operating since August 2010. Regardless of whether *Flame* did any genuine harm to Iran's oil and gas production, energy sector in Iran had suffered because of these cyber attacks [35].

A self-replicating virus (*Shamoon*) infected more than 30,000 devices of Saudi Aramco on 15 August 2012 which caused huge interruption to the world's biggest oil producer [36].

Another example of global cyberwar is *Turla* malware. This time former Eastern Bloc countries' diplomatic embassies were targeted. The main motivation behind the attacks was to monitor these embassies closely [37].

Kaspersky and Symantec both reported interestingly the revelation of a digital weapon framework which they called *Regin*. By, the malware had as of now been available for use for a long time and targeting many nations such as Germany, Belgium, Brazil, India and Indonesia. Some accuses *the Five Eyes Alliance*, which includes the US, Britain, Canada, Australia and New Zealand as a creator of the malware [38].

*IRATEMONK* gives programming application ingenuity on PCs embedding the hard drive firmware to gain execution through Master Boot Record (MBR) substitution which was allegedly created by *the Equation Group* who are assumed to be the most sophisticated and advanced cyber attack group in the world. Tens of thousands of victims were affected by the Equation Group's CNE (computer network exploitation) operations, including sectors like government and diplomatic institutions, telecoms, aerospace, energy, nuclear research, oil and gas, military, nanotechnology, Islamic activists and scholars, mass media, transportation, financial institutions, and Companies developing encryption technologies. They use exceptional and complex tools which they called their trojans, such as *EQUATIONLASER, EQUATIONDRUG, DOUBLEFANTASY,TRIPLEFANTASY, FANNY, GRAYFISH* and many others [39] [40] [41].

## V. CYBER WAREFARE LAW

It is important to know what kind of legal actions states can take globally when they are exposed to such cyber attacks. As said before, cyberspace is a new warfare and there are still so many legal loopholes in international regulations about cyber attacks. Former NSA Director Lt. Gen. Keith B. Alexander emphasized to the members of the Senate Armed Services Committee in 2010 that cyber warfare was advancing so quickly and there was a "mismatch between our technical capabilities to conduct operations and the governing laws and policies." Professor of Law and the Director of the Center for Terrorism Law at St. Mary's University School of Law Jeffrey F. Addicott has similar perspective. In his opinion "international laws associated with the use of force are woefully inadequate in terms of addressing the threat of cyber warfare [42] [43]."

According to many countries cyberwar is not any different than conventional war from angle of legal aspects. Cyberwar is within the scope of international legal system for war. Based on Article 5 of the North Atlantic Treaty, an attack on one Ally shall be considered an attack on all Allies. Recently, this article was extended by NATO members including cyber attacks [44]. Additionally, NATO Cooperative Cyber Defence Centre of Excellence which is a NATO-accredited research and training facility dealing with education, consultation, lessons learned, research and development in the field of cybersecurity was founded on 14 May, 2008 in Tallinn, Estonia [45].

There are two recognizable methods for taking a gander at war under international law. "Both of the two traditional branches of the law of war: (i) the *Jus ad bellum*, which governs resort to war, and (ii) the *Jus in bello*, which governs the conduct of hostilities [46]." These methods were applied for cyberwar as well. Previous experiences showed that preventing violent actions before happening is more critical and beneficial than trying to fix it after, which is valid for cyber conflicts too. To achieve this purpose the Law of Armed Conflict was created (LOAC). Generally the applicability of the LOAC regularly relied on a State subjectively characterizing a contention as a war. Recognition of a condition of war is no more required to trigger the LOAC. After the 1949 Geneva Conventions, the LOAC is currently activated by the presence of armed conflict between States. Even though the law is for armed conflicts, it can be a reference point for cyber warfare [47] [48]. 2(4) of Charter of the United Nations and Statute of International Court of Justice which was signed on 26 June 1945, in San Francisco clearly expresses that "All Members shall refrain in their international relations from the threat or use of force against the territorial integrity or political independence of any state, or in any other manner inconsistent with the Purposes of the United Nations [49]." Using armed attacks are only acceptable under the conditions of self-defense and agreement of UN Security Council. Article 51 explains that as "Nothing in the present Charter shall impair the inherent right of individual or collective self-defense if an armed attack occurs against a Member of the United Nations, until the Security Council has taken the measures necessary to maintain international peace and security. Measures taken by Members in the exercise of this right of self-defense shall be immediately reported to the Security Council and shall not in any way affect the authority and responsibility of the Security Council under the present Charter to take at any time such action as it deems necessary in order to maintain or restore international peace and security [49]." The right of self-defense is given to all members if there is a threat or use of force against the territorial integrity or political independence in case of armed attacks. Cyber attacks, on the other hand, aren't armed attacks for the UN therefore it can't be countered using armed assaults. A more common starting point for analysis is to consider the impacts or outcomes of a digital assault figuring out if it crosses the edge of armed attack. It was suggested that to qualify a cyber attack as an armed attack there must be violent consequences like bombs produce [50].

Geneva Conventions Article 49, however, defines attack as "... acts of violence against the adversary, whether in offence or in defence [51]." Instead of focusing armed attacks, Article 49 describes the attack based on its results. So if cyber attacks cause any violent outcomes, some legal experts say it should be considered within the scope of Charter of the United Nations Article 51. Aforementioned article does not apply to individuals or groups. It only applies to states. That means a person or a group can't be the subject of Article 51's self-defense. In this case, it is almost impossible to hold any state responsible for cyber attacks as long as it is admitted officially by them. For instance, although within the knowledge of everyone that Russian government sponsored digital attacks against Estonia in 2007, no one put the blame on Russian government formally since there isn't any concrete evidence. Moreover Responsibility of States for Internationally Wrongful Acts states that "... the basic rules of international law concerning the responsibility of States for their internationally wrongful acts. The emphasis is on the secondary rules of State responsibility: that is to say, the general conditions under international law for the State to be considered responsible for wrongful actions or omissions, and the legal consequences which flow therefrom. The articles do not attempt to define the content of the international obligations, the breach of which gives rise to responsibility. This is the function of the primary rules, whose codification would involve restating most of substantive customary and conventional international law [52]. International Humanitarian Law (IHL), also known as the laws of war is the structure for situations of armed conflict and occupation. The main purpose behind the law is limiting the effects of armed conflict mostly for civilians who are not directly part of the war [53]. Even supposing state level cyber weapons are intended to be used for military actions notwithstanding civilians can be harmed as well because of the consequences of these cyber attacks. It is also not clear whether IHL is applicable or not in this situations. For this purpose, International Cyber Incidents: Legal Considerations was published by Cooperative Cyber Defence Centre of Excellence (CCD COE) in 2010 to enlighten some issues about legal cyber challenges that urges legal authorities "To establish a robust and efficient cyber defence regime, legal and policy frameworks must have a multidisciplinary approach that incorporates legal jurisdictions for prevention and response to all-hazard threats; international collaboration and cooperation; an understanding of private sector legal rights and responsibilities (under public ordering such as regulations, as well as private ordering such as contracts) for the ICT systems and assets that are the new components of national and international security; and regard for the community of users [54]."

"In 2009, the NATO Cooperative Cyber Defence Centre of Excellence (NATO CCD COE), an international military organization based in Tallinn, Estonia, and accredited in 2008 by NATO as a 'Centre of Excellence', invited an independent 'International Group of Experts' to produce a manual on the law governing cyber warfare ... 'Tallinn Manual', results from an expert-driven process designated to produce a non-binding document applying existing law to cyber warfare." The most extensive study about cyber warfare legal concept is 'Tallinn Manual on the International Law Applicable to Cyber Warfare' even though it isn't a binding agreement. More than 30 experts from all over the world whose profession is in international law among with authorities from the United States Naval War College, the United Kingdom Royal Air Force, the Canadian Forces, the Swedish National Defence College, the University of Amsterdam, Chatham House, the Geneva Centre for Security Policy contributed the manual. It was prepared under the observatory of NATO, U.S. Cyber Command and International Committee of the Red Cross [55].

The second edition of Tallinn Manual (Tallinn 2.0) was published in 2016 [55].

The *jus ad bellum* (the law governing the use of force) and *jus in bello* (international humanitarian law) are the essential core interests of Tallinn Manual.

Tallinn Manual consists of two parts and 95 rules. International cybersecurity law and the law of cyber armed conflict are the titles of the parts. The first part has subtitles like state and cyberspace, the use of force, while second part's subtitles are the law of armed conflict generally, conduct of hostilities, certain persons, objects and activities, occupation and neutrality.

There are five rules under Section 1: Sovereignty, jurisdiction, and control. First 5 rules tries to draw a line for cyberspace based on cyber infrastructure within states' territories and explains the idea of sovereignty and jurisdiction.

*(Rule 1) A State may exercise control over cyber infrastructure and activities within its sovereign territory.*

Rules 6, 7, 8 and 9 slightly mention the law of State responsibility and precautions.

*(Rule 6) A State bears international legal responsibility for a cyber operation attributable to it and which constitutes a breach of an international obligation.*

*(Rule 9) A State injured by an internationally wrongful act resort to proportionate countermeasures, including cyber countermeasures, against the responsible State.*

It is hard to identify cyber armed attack so does self-defence against it. Armed attack was consistently seen as a higher limit than use of force. According to the manual "... the most disruptive and destructive cyber operations – those that qualify as 'armed attacks' and therefore allow States to respond in self-defence."

(Rule 13) A State that is the target of a cyber operation that rises to level of an armed attack may exercise its inherent right of self-defence. Whether a cyber operation constitutes an armed attack depends on its scale and effects.

One another significant issues that were addressed in Tallinn Manual are defining cyber attacks and protecting civilians. "In the Tallinn Manual, attacks include operations that cause injury or death to people or damage or destroy objects (Rule 30); any attack directed against civilians or civilian objects with these consequences is unlawful (Rules 31-32). Some experts stretched the *cyber attack* notion to include a cyber operation that engenders loss of functionality and thereby requires repair of the system [56]."

*(Rule 30) A cyber attack is a cyber operation, whether offensive or defensive, that is reasonable expected to cause injury or death or persons or damage or destruction to objects.*

*(Rule 31) The principle of distinction applies to cyber attacks.*

*(Rule 32) The civilian population as such, as well as individual civilians, shall not be the object of cyber attack.*

CONCLUSION

On April 1, 2015 President Barack Obama signed an executive order authorizing the Treasury Department to financially sanction anyone using cyber attacks "that create a significant threat to the national security, foreign policy or economic health or financial stability of the United States [57]." Taking legal countermeasures against cyber attacks is being more important and serious. Despite the fact that apparently there is confusion about international legal actions for cyber warfare. Either in state-level or individual, hackers should be put on trial due process of law. Admiral Michael Rogers who is the current Commander of U.S. Cyber Command and Director of the National Security Agency says "Remember, anything we do in the cyber arena ... must follow the law of conflict. Our response must be proportional, must be in line with the broader set of norms that we've created over time. I don't expect cyber to be any different [58]." Vulnerability of internet will continue for quite a while accordingly, it is almost impossible to avoid all those cyber attacks, espionage and sabotage. That's why it is so critical to know for the states which were exposed cyber attacks their legal rights and how to respond them legally to prevent chaos or maybe even something more. Even though, there are some guidance about how law should applied in case of cyber attacks/war, it is simply not enough and not obligated by the international authorities. Cyber warfare law is the new area of research for international laws and especially for international war law.

ACKNOWLEDGMENT

This research was partially supported by the Scientific and Technological Research Council of Turkey (TUBITAK) Informatics and Information Security Research Center (BILGEM) and Istanbul Sehir University. We thank our colleagues from TUBITAK BILGEM and our lecturer from Istanbul Sehir University who provided insight and expertise that greatly assisted the research, although they may not agree with all of the interpretations/conclusions of this paper.

BIBLIOGRAPHY

[1] The White House, "the National Strategy to Secure Cyberspace," the White House, Ed. 2003. [Online]. Available: https://www.us-cert.gov/sites/default/files/publications/cyberspace_strategy.pdf. Accessed: Feb. 4, 2016. D. [30]

[2] U.S. DoD, "Department of Defense Dictionary of Military and Associated Terms," U.S. Department of Defense, 2014. [Online]. Available:

http://www.dtic.mil/doctrine/new_pubs/jp1_02.pdf. Accessed: Feb. 4, 2016.

[3] NATO Cooperative Cyber Defence Centre of Excellence, "National Cyber Security Framework Manual," 2012. [Online]. Available: https://ccdcoe.org/publications/books/NationalCyberSecurityFrameworkManual.pdf. Accessed: Feb. 4, 2016.

[4] J. A. Lewis, "Thresholds for Cyberwar," Center for Strategic and International Studies, 2010. [Online]. Available: http://csis.org/files/publication/101001_ieee_insert.pdf. Accessed: Feb. 5, 2016.

[5] F. Schreier, "On Cyberwarfare," Geneva Centre for the Democratic Control of Armed Forces, 2012. [Online]. Available: http://www.dcaf.ch/content/download/67316/1025687/file/OnCyberwarfare-Schreier.pdf. Accessed: Feb. 5, 2016.

[6] RAND Corporation, "Cyber Warfare," 2016. [Online]. Available: http://www.rand.org/topics/cyber-warfare.html.

[7] R. A. Clarke and R. K. Knake, *Cyber War: The Next Threat to National Security and What to do About it*. New York: HarperCollins Publishers, 2010.

[8] U.S. DoD, *Cyber Command Fact Sheet* in *Cyber Command Fact Sheet*. U.S. Department of Defense, 2010. [Online]. Available: http://www.stratcom.mil/factsheets/cc.

[9] NSA, *NSA/CSS Strategy*. National Security Agency, 2010. [Online]. Available: https://www.nsa.gov/about/_files/nsacss_strategy.pdf. Accessed: Feb. 8, 2016.

[10] US-CERT, "About us," in *United States Computer Emergency Readiness Team*. [Online]. Available: https://www.us-cert.gov/about-us. Accessed: Feb. 8, 2016.

[11] FBI, "Addressing Threats to the Nations Cybersecurity," Federal Bureau of Investigation, 2011. [Online]. Available: https://www.fbi.gov/about-us/investigate/cyber/addressing-threats-to-the-nations-cybersecurity-1. Accessed: Feb. 8, 2016.

[12] CNCERT/CC, "Weekly Report," in "Weekly Report," The National Computer Network Emergency Response Technical Team/Coordination Center of China, 2016.

[13] Ministry of National Defense of the People's Republic of China, "Cyber Warfare," 2012. [Online]. Available: http://www.mod.gov.cn/affair/2012-03/29/content_4354898.htm. Accessed: Feb. 8, 2016.

[14] D. Ball, "China's Cyber Warfare Capabilities," *Security Challenges*, vol. 7, no. 2, pp. 81–103, 2011. [Online]. Available: http://www.informa.com.au/download/cyber-security-summit-speaker-presentations/Desmond%20Ball%20Paper.pdf. Accessed: Feb. 8, 2016.

[15] Z. Tao, "China's First 'Blue Army': Powerful Rivals," in *Ministry of National Defense The People's Republic of China*, 2015. [Online]. Available: http://eng.mod.gov.cn/Opinion/2015-06/05/content_4588571.htm. Accessed: Feb. 8, 2016.

[16] The Russian Government, "Federal Security Service," in *the Russian Government*, 2002. [Online]. Available: http://government.ru/en/department/113/. Accessed: Feb. 9, 2016.

[17] Federal Security Service, "Federal Security Service," in *Federal Security Service*. [Online]. Available: http://www.fsb.ru/fsb. Accessed: Feb. 9, 2016.

[18] Spetssvyaz, "The Special Communications and Information Service of the Federal Protective Service of the Russian Federation," in *the Special Communications and Information Service of the Federal Protective Service of the Russian Federation*. [Online]. Available: http://www.fso.gov.ru/struktura/p2_1_1.html. Accessed: Feb. 9, 2016.

[19] K. Giles, "Russian Cyber Security: Concepts and Current Activity," Chatham House, 2012. [Online]. Available: https://www.chathamhouse.org/sites/files/chathamhouse/public/Research/Russia%20and%20Eurasia/060912summary.pdf. Accessed: Feb. 9, 2016.

[20] J. Valentino DeVries and D. Yadron, "Cataloging the World's Cyberforces," in *The Wall Street Journal*, wsj.com, 2015. [Online]. Available: http://www.wsj.com/articles/cataloging-the-worlds-cyberforces-1444610710. Accessed: Feb. 9, 2016.

[21] W. Carroll, "Inside DPRK's unit 121," in *Defensetech*, 2007. [Online]. Available: http://defensetech.org/2007/12/24/inside-dprks-unit-121/. Accessed: Feb. 9, 2016.

[22] V. D. Cha, "What do they really want? Obama's North Korea conundrum," *The Washington Quarterly*, vol. 32, no. 4, pp. 119–138, Oct. 2009.

[23] NATO Review Magazine, "Cyber Timeline," in *North Atlantic Treaty Organization*. [Online]. Available: http://www.nato.int/docu/review/2013/cyber/timeline/EN/index.htm. Accessed: Feb. 9, 2016.

[24] R. Langner, "Stuxnet: Dissecting a Cyberwarfare Weapon," in *IEEEXplorre*, vol. 9, IEEE, 2011, pp. 49–51. [Online]. Available: http://ieeexplore.ieee.org/stamp/stamp.jsp?tp=&arnumber=5772960. Accessed: Feb. 9, 2016.

[25] J. P. Farwell and R. Rohozinski, "Stuxnet and the Future of Cyber war," *Survival*, vol. 53, no. 1, pp. 23–40, Feb. 2011.

[26] N. Falliere, L. O. Murchu, and E. Chien, "W32.Stuxnet Dossier," Symantec, 2011. [Online]. Available: https://www.symantec.com/content/en/us/enterprise/media/security_response/whitepapers/w32_stuxnet_dossier.pdf. Accessed: Feb. 9, 2016.

[27] D. Kushner, "The Real Story of Stuxnet," in *IEEE Spectrum*, 2013. [Online]. Available: http://spectrum.ieee.org/telecom/security/the-real-story-of-stuxnet. Accessed: Feb. 9, 2016.

[28] S. J. Shackelford, "From Nuclear War to Net War: Analogizing Cyber Attacks in International Law," 2009.

[29] S. Herzog, "Revisiting the Estonian Cyber Attacks: Digital Threats and Multinational Responses," *Journal of Strategic Security*, vol. 4, no. 2, pp. 49–60, 2015.

[30] D. Hollis, "Cyberwar Case Study: Georgia 2008," in *Small War Journal*, 2011.

[31] J. A. Lewis, "Computer Espionage, Titan Rain and China," Center for Strategic and International Studies, 2005.

[32] C. Tankard, "Advanced Persistent Threats and how to Monitor and Deter Them," *Network Security*, vol. 2011, no. 8, pp. 16–19, Aug. 2011. [Online]. Available: http://www.sciencedirect.com/science/article/pii/S1353485811700861. Accessed: Feb. 10, 2016.

[33] Google Official Blog, "A New Approach to China," Official Google Blog, 2010. [Online]. Available: https://googleblog.blogspot.com.tr/2010/01/new-approach-to-china.html. Accessed: Feb. 10, 2016.

[34] Kaspersky Lab, "Kaspersky Lab Identifies Operation 'Red October,' an Advanced Cyber-Espionage Campaign Targeting Diplomatic and Government Institutions Worldwide," in *Kaspersky Lab*, 2013.

[35] D. Lee, "Flame: Massive Cyber-attack Discovered, Researchers Say," in *BBC Technology*, BBC News, 2012. [Online]. Available: http://www.bbc.com/news/technology-18238326. Accessed: Feb. 10, 2016.

[36] C. Bronk and E. Tikk-Ringas, "The Cyber Attack on Saudi Aramco," *Survival*, vol. 55, no. 2, pp. 81–96, May 2013.

[37] E. Hanford, "The Cold War of Cyber Espionage," in *Heinonline*, 2014. [Online]. Available: http://heinonline.org/HOL/Page?handle=hein.journals/pilr20&div=9&g_sent=1&collection=journals. Accessed: Feb. 10, 2016.

[38] Spiegel Online, "Source Code Similarities: Experts Unmask 'Regin' Trojan as NSA Tool," in *Spiegel Online*, SPIEGEL ONLINE, 2015.

[39] F-Secure Labs, "The Equation Group Equals NSA / IRATEMONK," in *F-Secure Labs*, 2015. [Online]. Available: https://www.f-secure.com/weblog/archives/00002791.html. Accessed: Feb. 10, 2016.

[40] Kaspersky Lab's Global Research and Analysis Team (GReAT), "Equation Group: Questions and Answers," Kaspersky Lab's, 2015. [Online]. Available: https://securelist.com/files/2015/02/Equation_group_questions_and_answers.pdf. Accessed: Feb. 10, 2016.

[41] Kaspersky Lab's Global Research & Analysis Team (GReAT), "Equation: The Death Star of Malware Galaxy," 2015. [Online]. Available: https://securelist.com/blog/research/68750/equation-the-death-star-of-malware-galaxy/. Accessed: Feb. 10, 2016.

[42] T. Shanker, "Cyberwar Nominee Sees Gaps in Law," in *The New York Times*, The New York Times, 2014. [Online]. Available: http://www.nytimes.com/2010/04/15/world/15military.html. Accessed: Feb. 10, 2016.

[43] C. J. Dunlap, "Perspectives for Cyber Strategists on Law for Cyberwar," Strategic Studies Quarterly, 2011. [Online]. Available: http://scholarship.law.duke.edu/cgi/viewcontent.cgi?article=2992&context=faculty_scholarship. Accessed: Feb. 10, 2016.

[44] North Atlantic Treaty Organization, "Collective Defence - Article 5," in *North Atlantic Treaty Organization*, 2015. [Online]. Available: http://www.nato.int/cps/en/natohq/topics_110496.htm? Accessed: Feb. 10, 2016.

[45] R. Czulda and R. Los, *NATO Towards the Challenges of a Contemporary World: 2013*. Warsaw: International Relations Research Institute in Warsaw, 2013.

[46] R. D. Sloane, "The Cost of Conflation: Preserving the Dualism of Jus ad Bellum and Jus in Bello in the Contemporary Law of War," in *Social Science Research Network*, 2008.

[47] International and Operational Law Department, The United States Army Judge Advocate General's Legal Center and School, "Law of Armed Conflict Deskbook," The United States Army Judge Advocate Genral's Legal Center and School, 2012.

[48] UN General Assembly Security Council, "Prevention of Armed Conflict - Report of the Secretary-General," United Nations, 2001. [Online]. Available: http://unpan1.un.org/intradoc/groups/public/documents/un/unpan005902.pdf. Accessed: Feb. 10, 2016.

[49] United Nations, "Charter of the United Nations and Statute of International Court of Justice," United Nations, 1945. [Online]. Available: https://treaties.un.org/doc/Publication/CTC/uncharter.pdf. Accessed: Feb. 10, 2016.

[50] M. C. Waxman, "Self-defensive Force Against Cyber Attacks: Legal, Strategic and Political Dimensions," in *Social Science Research Network*, 2013. [Online]. Available: http://papers.ssrn.com/sol3/papers.cfm?abstract_id=2235838. Accessed: Feb. 10, 2016.

[51] United Nations, "Protocol Additional to the Geneva Conventions of 12 August 1949," United Nations, 1977. [Online]. Available: https://treaties.un.org/doc/Publication/UNTS/Volume%201125/volume-1125-I-17512-English.pdf. Accessed: Feb. 10, 2016.

[52] United Nations, "Draft Articles on Responsibility of States for internationally Wrongful Acts, with Commentaries," in United Nations, 2001. [Online]. Available: http://legal.un.org/ilc/texts/instruments/english/commentaries/9_6_2001.pdf. Accessed: Feb. 10, 2016

[53] International Justice Research Center, "International Humanitarian Law," in International Justice Research Center, International Justice Resource Center, 2012. [Online]. Available: http://www.ijrcenter.org/international-humanitarian-law/. Accessed: Feb. 10, 2016.

[54] E. Tikk, K. Kaska, and L. Vihul, *International Cyber Incidents: Legal Considerations*. Cooperative Cyber Defence Centre of Excellence (CCD COE), 2010. [Online]. Available: https://ccdcoe.org/publications/books/legalconsiderations.pdf. Accessed: Feb. 10, 2016.

[55] NATO Cooperative Cyber Defence Centre of Excellence, *Tallinn manual on the international law applicable to cyber warfare: Prepared by the international group of experts at the invitation of the NATO cooperative Cyber Defence centre of excellence*, M. N. Schmitt, Ed. Cambridge: Cambridge University Press, 2013.

[56] L. Vihul, "The Tallinn Manual on the international Law Applicable to Cyber Warfare," in *The European Journal of International Law*, EJIL: Talk!, 2013. [Online]. Available: http://www.ejiltalk.org/the-tallinn-manual-on-the-international-law-applicable-to-cyber-warfare/. Accessed: Feb. 11, 2016.

[57] C. Spodak, "Obama Announces Executive Order on Sanctions Against Hackers," in *CNN*, CNN, 2015. [Online]. Available: http://edition.cnn.com/2015/04/01/politics/obama-cyber-hackers-executive-order/. Accessed: Feb. 11, 2016.

[58] P. Tucker, "NSA Chief: Rules of War apply to Cyberwar, Too," in *Defense One*, Defense One, 2015. [Online]. Available: http://www.defenseone.com/technology/2015/04/nsa-chief-rules-war-apply-cyberwar-too/110572/. Accessed: Feb. 11, 2016.